\documentclass[dvips]{tlp}

\usepackage{alltt}
\usepackage{graphics}
\usepackage{html}

\renewcommand{\_}{\char'137}
\newcommand{\figpath}{.}

\newcommand{\wlib}
 {\latex{\mbox{\textsf{P}\textit{i}\textsf{LL}\textit{o}\textsf{W}}}%
  \html{\mbox{PiLLoW}}}

\newenvironment{descpredicates}{\begin{description}}{\end{description}}
\newcommand{\predicateitem}[1]{\item {\texttt{#1}}}
\newcommand{\var}[1]{\textsl{#1}}
\newcommand{\ttvar}[1]{\texttt{\textsl{#1}}}

\title[Distributed WWW Programming using (Ciao-)Prolog and the \wlib\
 lib] 
{Distributed WWW Programming using (Ciao-)Prolog and the \wlib\
 library\footnotemark} 

\author[D. Cabeza and M. Hermenegildo]
       {DANIEL CABEZA and MANUEL HERMENEGILDO\\
        \htmladdnormallink{CLIP Group}{http://www.clip.dia.fi.upm.es}\\
        \htmladdnormallink{\texttt{http://www.clip.dia.fi.upm.es} -- \texttt{http://www.cliplab.org} }{http://www.clip.dia.fi.upm.es}\\
        Facultad de Inform\'{a}tica\\
        Universidad Polit\'{e}cnica de Madrid (UPM)\\
        28660-Boadilla del Monte, Madrid, Spain\\
        \email{\{dcabeza,herme\}@fi.upm.es}}

\volume{1(3)}
\pagerange{251--282}
\pubyear{2001}
\jdate{May 2001}

\begin{document}

\maketitle

\begin{htmlonly}
\bodytext{BGCOLOR="#ffffff" TEXT="000000"} 
\end{htmlonly}

\footnotetext{This paper is an expanded and improved version of
    \cite{html-pl-www,html-pl-complangprolog-www,pillow-www6,pillow-ws}.}

\begin{abstract}
  We discuss from a practical point of view a number of issues
  involved in writing distributed Internet and WWW applications using
  LP/CLP systems.  We describe \wlib, a \emph{public-domain} Internet
  and WWW programming library for LP/CLP systems that we have designed
  in order to simplify the process of writing such applications.
  \wlib\ provides facilities for accessing documents and code on the
  WWW; parsing, manipulating and generating HTML and XML structured
  documents and data; producing HTML forms; writing form handlers and
  CGI-scripts; and processing HTML/XML templates.  An important
  contribution of \wlib\ is to model HTML/XML code (and, thus, the
  content of WWW pages) as terms.  The \wlib\ library has been
  developed in the context of the Ciao Prolog system, but it has been
  adapted to a number of popular LP/CLP systems, supporting most of
  its functionality. We also describe the use of concurrency and 
  a high-level model of client-server interaction, Ciao Prolog's {\em
    active modules}, in the context of WWW programming.  We propose a
  solution for client-side downloading and execution of Prolog code,
  using generic browsers.  Finally, we also provide an overview of
  related work on the topic.

\end{abstract}

\section{Introduction}

The wide diffusion of the Internet and the popularity of the ``World
Wide Web'' \cite{www} --WWW-- protocols are effectively providing a
novel platform that facilitates the development of new classes of
portable and user-friendly distributed applications.  Good support for
network connectivity and the protocols and communication architectures
of this novel platform are obviously requirements for any programming
tool to be useful in this arena. However, this alone may not be
enough. It seems natural that significant parts of network
applications will require symbolic and numeric capabilities which are
not necessarily related with distribution. Important such capabilities
are, for example, high-level symbolic information processing, dealing
with combinatorial problems, and natural language processing in
general.  Logic Programming (LP) \cite{kowalski74,Colmerauer75} and
Constraint Logic Programming (CLP) systems
\cite{Jaff87,pascal-book,colmerauer90,dinc90,eclipse} have been shown
particularly successful at tackling these issues (see, for example,
the proceedings of recent conferences on the ``Practical Applications
of Prolog'' and ``Practical Applications of Constraint Technology'').
It seems natural to study how LP/CLP technology fares in developing
applications which have to operate over the Internet.

In fact, Prolog, its concurrent and constraint based extensions, and
logic programming languages in general have many characteristics which
appear to set them particularly well placed for making an impact on
the development of practical networked applications, ranging from the
simple to the quite sophisticated.  Notably, LP/CLP systems share many
characteristics with other recently proposed network programming tools,
such as Java, including dynamic memory management, well-behaved
structure and pointer manipulation, robustness, and compilation to
architecture-independent bytecode. Furthermore, and unlike the
scripting or application languages currently being proposed (e.g.,
shell scripts, Perl, Java, etc.), LP/CLP systems offer a quite unique
set of additional features including dynamic databases, search
facilities, grammars, sophisticated meta-programming, and well
understood semantics. 

In addition, most LP/CLP systems also already offer some kind of low
level support for remote communication using Internet protocols. This
generally involves providing a {\em sockets} (ports) interface whereby
it is possible to make remote data connections via the Internet's
native protocol, TCP/IP. A few systems support higher-level
communication layers on top of this interface including linda-style
blackboards (e.g., SICStus Prolog \cite{sicstus}, Ciao
\cite{shared-database,ciao-dis-impl-parimp-www,ciao-ppcp,ciao-ilps95,ciao-novascience,prog-glob-an,ciao-reference-manual-tr},
BinProlog/$\mu^2$-Prolog
\cite{binprolog-complangprolog-www,multi-prolog}, etc.) or shared
variable-based communication (e.g., KL1 \cite{klic-evanws}, AKL
\cite{ilps-akl-paradigms}, Oz \cite{kernel-oz-94}, Ciao
\cite{att-var-iclp,ciao-dis-impl-parimp-www}, etc.).  In some cases,
this functionality is provided via libraries, building on top of the
basic TCP/IP primitives. This is the case, for example, of the SICStus
and Ciao distributed linda-style interfaces. In fact, as we have shown
in previous work, shared-variable based communication can also be
implemented in conventional systems via library predicates, by using
attributed variables \cite{att-var-iclp,ciao-dis-impl-parimp-www}. In
addition to these communication primitives, several 
systems offer concurrency and even higher-level abstractions
(distributed objects, mobile code, ...) which are very useful for
developing general-purpose distributed applications. 

Our concrete interest here is WWW applications. These applications
generally use specific high-level protocols (such as {\tt HTTP} or
{\tt FTP}), data formats (such as {\tt HTML} or {\tt XML}), and
application architectures (e.g., the CGI interface) which are
different from, e.g., the shared-variable or linda-based protocols
typically used in other types of distributed applications. In this
paper we study how good support for these WWW-related protocols, data
formats, and architectures can be provided for LP/CLP systems,
building on the widely available interfaces to the basic TCP/IP
protocols.  Our aim is to discuss from a practical point of view a
number of the new issues involved in writing WWW applications using
LP/CLP systems, as well as the architecture of some typical solutions.
In the process, we will describe \wlib\ (``Programming in Logic
Languages on the Web''), a public domain Internet/WWW programming
library for LP/CLP systems which, we argue, significantly simplifies
the process of writing such applications. \wlib\ provides facilities
for generating HTML/XML structured documents by handling them as Herbrand
terms, producing HTML {\em forms}, writing form handlers, processing
HTML/XML templates, accessing and parsing WWW documents, etc.
 We also describe the architecture of some relatively
sophisticated application classes, using a high-level model of
client-server interaction, {\em active modules}
\cite{ciao-dis-impl-parimp-www}.  Finally we describe an architecture
for automatic LP/CLP code downloading for local execution, using just
the library and generic browsers.

Apart from the tutorial value of the paper, we present a number of
technical contributions which include the idea of representing HTML
and XML code (and structured text in general) as Prolog terms, the use
of the logical variable in such terms leading to a model of an ``HTML
template'' (a pair comprising a term with free variables and a
dictionary associating names to those variables), the notion of
``active logic modules'' and its application to solving efficiency
issues in CGI interaction in a very simple way, the idea of ``Prolog
scripts'' and its application to CGIs, and the identification of a
number of features that should be added to existing systems in order
to facilitate the programming of WWW applications --mainly
concurrency.

The argument throughout the paper is that, with only very small
limitations in functionality (which disappear when concurrency is
added, as in systems such as BinProlog/$\mu^2$-Prolog, AKL, Oz, KL1,
and Ciao Prolog), it is possible to add an extremely useful
Internet/WWW programming layer to any LP/CLP system without making any
significant changes in the implementation. We argue that this layer
can simplify the generation of applications in LP/CLP systems
including active WWW pages, search tools, content analyzers, indexers,
software demonstrators, collaborative work systems, MUDs and MOOs,
code distributors, etc.

The purpose of the paper is also to serve as
a tutorial,
containing sufficient information for developing relatively complex
WWW applications in Prolog and other LP and CLP languages using the
\wlib\ library.  The \wlib\ library has been developed in the context
of the Ciao Prolog system, but it has been adapted to a number of
other popular LP/CLP systems, supporting most of its functionality.
The Ciao Prolog system and the \wlib\ library can be freely downloaded
from
\htmladdnormallink{\texttt{http://www.clip.dia.fi.upm.es}}{http://www.clip.dia.fi.upm.es}
and
\htmladdnormallink{\texttt{http://www.cliplab.org}}{http://www.cliplab.org}.

\section{Writing basic cgi-bin applications}

\begin{figure}
\centerline{\includegraphics{\figpath/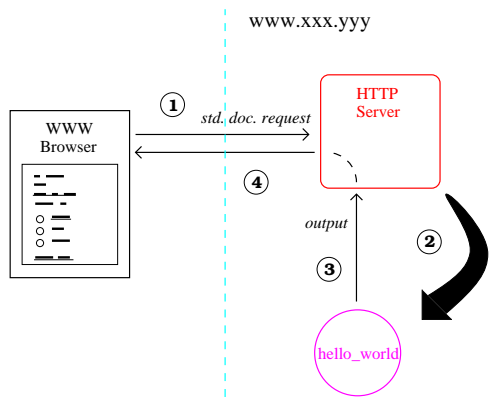}}
\caption{The CGI interface}
\label{fig:cgi}
\end{figure}

The simplest way of writing WWW applications is through the use of the
``Common Gateway Interface'' (CGI).  A CGI executable is a standard
executable file but such that the HTTP server (the program that
responds to HTTP requests in a machine which serves a WWW site) can
tell that it in fact contains a program that is to be run, rather than
a document text that is to be sent to the client (the browser) as
usual. The file can be distinguished by belonging to a special
directory, commonly named {\tt cgi-bin}, or by a special filename
ending, such as {\tt .cgi}.  This is normally set during configuration
of the HTTP server. The basic idea behind the CGI interface is
illustrated in Figure \ref{fig:cgi}.  When the user selects an address
of a CGI executable in a document, such as {\tt
  http://www.xxx.yyy/cgi-bin/hello\_world} (or perhaps {\tt
  http://www.xxx.yyy/foo/hello\_world.cgi}) the browser issues a
standard document request (1). The {\tt HTTP} server, recognizing that
it is a CGI executable rather than a document, starts the executable
(2), and during such execution stores the output of the executable in
a buffer (3).  Upon termination of the executable, the contents of the
buffer (which should be in a format that the browser can handle, such
as HTML) are returned to the browser as if a normal page with that
content had been accessed (4).

The following is an example of how a very simple such executable can
be written in an LP/CLP language. The source might be as
follows:\footnote{Note that in the examples presented, and in order to
  shorten them, the HTML code may be slightly simplified, and as a result
  of this it may not be completely standard-conforming. However, the
  examples can be used as is with all popular browsers.} 

\begin{verbatim}
main :-
    write('Content-type: text/html'), nl, nl,
    write('<HTML>'),
    write('Hello world.'),
    write('</HTML>').
\end{verbatim}

\noindent
And the actual executable could be generated as usual, for example in
the Ciao system, using the stand\-alone compiler, by writing at a
UNIX shell ``\texttt{ciaoc -o hello\_world.cgi
  hello\_world}''.\footnote{It often convenient to use options (such
  as \texttt{ciaoc}'s \texttt{-s} or \texttt{-S}) which will generate
  a standalone executable which is independent of any libraries.}  The
executable then has to be placed in an appropriate place (accessible
via an HTTP address by a browser) and have the right permissions for
being executed by the server (for example, in some systems this means
being executable by the user ``\texttt{nobody}'').

In systems which make executables through saved states
(which usually have the disadvantage of their generally large size),
at the system prompt one could create the executable by writing
something like:

\begin{verbatim}
?:- compile('hello_world.pl'), 
     save('hello_world.cgi'), main.
\end{verbatim}

\section{LP/CLP Scripts for CGI Applications}

CGI executables are often small- to medium-sized programs that perform 
relatively simple tasks. This, added to the slow speed of the
network connection in comparison with that of executing a program
(which makes program execution speed less important) has made
scripting languages (such as shell scripts or Perl) very popular for
writing these programs. The popularity is due to the fact that no
compilation is necessary (extensive string handling capabilities also
play an important role in the case of Perl), and thus changes and
updates to the program imply only editing the source file. 

Logic languages are, a priori, excellent candidates to be used as
scripting languages.\footnote{For example, the built-in grammars and databases
greatly simplify many typical script-based applications.} 
However, the relative complication in making executables (needing in
some systems to start the top-level, compile or consult the file, and
make a saved state) and the often large size of the resulting
executables may deter CGI application programmers.  It appears
convenient to provide a means for LP/CLP programs to be executable as
scripts, even if with reduced performance.

It is generally relatively easy to support scripts with the same
functionality in most LP/CLP systems.  In Ciao, the program {\tt
  ciao-shell} --which has also been adapted to SICStus
\cite{sicstus-scripts-complangprolog-www}-- accomplishes this task, by
first loading the file given to it as the first argument (but skipping
the first lines and avoiding loading messages) and then starting
execution at {\tt main/1} (the argument provides the list of command
line options). Then, for example, in a Unix system, the following
program can be run directly as a script without any need for
compilation:

\begin{verbatim}
#!/usr/local/bin/ciao-shell

main(_) :-
    write('Content-type: text/html'), nl, nl,
    write('<HTML>'),
    write('Hello world.'),
    write('</HTML>').
\end{verbatim}

Note that in some UNIX versions either the program \texttt{ciao-shell}
must be included in the \texttt{/etc/shells} listing or the first line
should be replaced by these two:
\begin{verbatim}
#!/bin/sh
exec ciao-shell $0 "$@"
\end{verbatim}

The execution of Prolog scripts may be optimized in some systems. For
example, in Ciao, the first time a script is run it is also compiled
and its bytecode is saved to a file.  At subsequent times, if the
script has not changed, the object code is retrieved from that file,
avoiding compilation or interpretation overhead.

\section{Form Handling in HTTP} 

\begin{figure}
\centerline{\includegraphics{\figpath/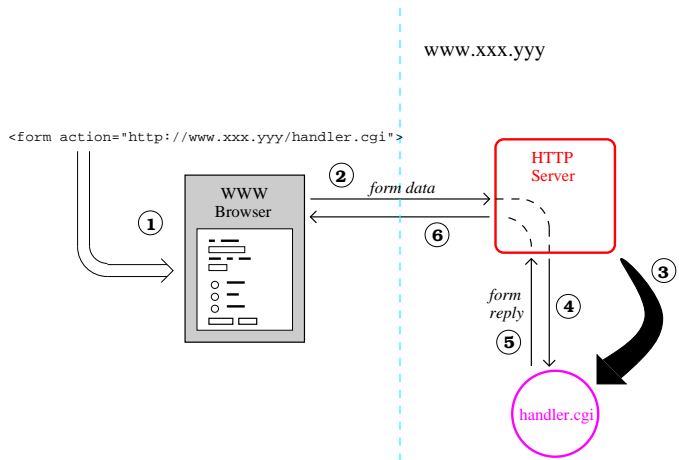}}
\caption{The Forms interface}
\label{fig:forms}
\end{figure}

So far we have shown CGI executables which produce output, but this
output is not a function of input coming from the request, which is
obviously of limited interest. CGI executables become most useful when
combined with HTML forms. HTML forms are HTML documents (or parts of
HTML documents) which include special fields such as text areas,
menus, radio buttons, etc. which allow providing input to CGI
executables.  The steps involved in the handling of the input
contained in a form are illustrated in Figure \ref{fig:forms}.  When a
document containing a form is accessed via a form-capable browser
(Mosaic, Netscape, Lynx, etc.), the browser displays the input fields,
buttons, menus, etc.\ indicated in the document, and {\em locally}
allows the user to perform input by modifying such fields.  However,
this input is not ultimately handled by the browser.  Instead, it will
be sent to a ``handler'' CGI program, which can be anywhere on the
net, and whose address must be given in the form itself (1). Forms
generally have a ``submit'' button such that, when pressed, the input
provided through the menus, text areas, etc.\ is sent by the browser
to the {\tt HTTP} server corresponding to the handler (2). Two methods
for sending this input exist: ``GET'' and ``POST''. In the meantime,
the sending browser waits for a response from that program, which
should come in the form of a new HTML document. The handler program is
invoked in much the same way as a cgi-bin application (3), except that
the information from the form is supplied to the handler (in different
ways depending on the system, the method of invocation, and the
content type) (4). This information is encoded in a predefined format,
which relates each piece of information to the corresponding field in
the form, by means of a keyword associated with each field.  The
handler then identifies the information corresponding to each field in
the original form, processes it, and then responds by writing an HTML
document to its standard output (5), which is forwarded by the server
to the waiting browser when the handler terminates (6).  An important
point to be noted is that, as with simple cgi-bin applications, the
handler is started and should terminate for each transaction. The
reader is referred, for example, to \cite{cgi-forms-tut-www}
for a more complete introduction to CGI scripts and HTML forms.

\section{Writing Form Handlers with \wlib}

The only complication in writing form handlers compared to writing
simple CGI applications is the need to capture and parse the form
data.  As we said before, this data can be provided in several ways,
depending on the system and the method used to invoke the form, and is
encoded with escape sequences. It is relatively easy to write a Prolog
program to parse such input (using, for example, definite clause
grammars --DCGs).  The \wlib\ library provides some predicates which
do this and simplify the whole task, hiding the low-level protocol
behind. The principal predicates provided include:

\begin{descpredicates}
  \predicateitem{get\_form\_input(\var{Dic})} Translates input from the
  form (with either the POST or GET methods, and even with
  \texttt{CONTENT\_TYPE} 
  multipart/form-data) to a dictionary \var{Dic} of
  \var{attribute}=\var{value} pairs. It translates empty \var{value}s
  (which indicate only the presence of an attribute) to the atom {\tt
    '\$empty'}, values with more than one line (from text areas or
  files) to a list of lines as strings, the rest to atoms or numbers
  (using \texttt{name/2}).  This is implemented using DCG parsers.

  \predicateitem{get\_form\_value(\var{Dic},\var{Var},\var{Val})} Gets
  value \var{Val} for attribute \var{Var} in dictionary \var{Dic}. Does
  not fail: value is {\tt ''} if not found (this simplifies merging form
  producers and form handlers, see later).

  \predicateitem{form\_empty\_value(\var{V})} Useful to check that a
  value \var{V} from a text area is empty (filters spaces, newlines, 
  linefeeds, etc.).

  \predicateitem{form\_default(\var{Val},\var{Default},\var{NewVal})}
  Useful when a form is only partially filled (and also in the first
  invocation of a combined form handler/producer -- see Section
  \ref{sec:merging}). If the value of \var{Val} is empty then
  \var{NewVal}=\var{Default}, else \var{NewVal}=\var{Val}.

  \predicateitem{my\_url(\var{URL})} Returns in \var{URL} the Uniform
  Resource Locator (WWW address) of this cgi executable.

  \predicateitem{form\_request\_method(\var{Method})} Returns in
  \var{Method} the method of invocation of the form handler (``{\tt
  GET}'' or ``{\tt POST}'').

\end{descpredicates}

For example, suppose we want to make a handler which implements a
database of telephone numbers and is queried by a form including a
single entry field with name \texttt{person\_name}. The handler might be
coded as follows:

\begin{verbatim}
#!/usr/local/bin/ciao-shell

:- include(library(pillow)).

main(_) :-
    get_form_input(Input),
    get_form_value(Input,person_name,Name),
    write('Content-type: text/html'), nl, nl,
    write('<HTML><TITLE>Telephone database</TITLE>'), nl,
    write('<IMG SRC="phone.gif">'),
    write('<H2>Telephone database</H2><HR>'),
    write_info(Name),
    write('</HTML>').

write_info(Name) :-
    form_empty_value(Name) ->
       write('You have to provide a name.')
  ; phone(Name, Phone) ->
       write('Telephone number of <B>'),
       write(Name),
       write('</B>: '),
       write(Phone)
  ; write('No telephone number available for <B>'),
    write(Name),
    write('</B>.').

phone(daniel, '336-7448').
phone(manuel, '336-7435').
phone(sacha,  '543-5316').
\end{verbatim}

The code above is quite simple. On the other hand, the interspersion
throughout the text of calls to \texttt{write} with HTML markup inside
makes the code somewhat inelegant. Also, there is no separation
between computation and input/output, as is normally desirable.  It
would be much preferable to have an encoding of HTML code as Prolog
terms, which could then be manipulated easily in a more elegant way,
and a predicate to translate such terms to HTML for output. This
functionality, provided by the \wlib\ library, is presented in the
next section.

\section{Handling HTML as Prolog Terms}
\label{sec:termconversion}

Since LP/CLP systems perform symbolic processing using Herbrand terms,
it seems natural to be able to handle HTML code directly as terms.
Then, such structures only need to be translated by appropriate
predicates to HTML code when they need to be output.  In general, this
relationship between HTML code and Prolog terms allows viewing any WWW
page as a Herbrand term.  The predicates which provide this
functionality in \wlib\ are:

\begin{descpredicates}
  \predicateitem{output\_html(\var{F})} Accepts in \var{F} an HTML term
  (or a list of HTML terms) and sends to the standard output the text
  which is the rendering of the term(s) in HTML format.
  
  \predicateitem{html2terms(\var{Chars},\var{Terms})} (also,
  \texttt{xml2terms/2}) Relates a list of HTML (resp.~XML) terms and a
  list of ASCII characters which are the rendering of the terms in
  HTML format. This predicate is reversible (but it normalizes in the
  reverse direction --see later). {\tt output\_html/2} uses this
  predicate to transform HTML terms in characters.  Again, this is
  implemented via DCG parsing.
\end{descpredicates}

In an {\em HTML term} certain atoms and structures represent special
functionality at the HTML level. An HTML term can be recursively a list
of HTML terms.  The following are legal HTML terms:

\begin{verbatim}
hello
[hello, world]
["This is an ", em('HTML'), " term"]
\end{verbatim}

When converting HTML terms to characters, {\tt html2terms/2} translates
special structures into the corresponding format in HTML, applying
itself recursively to their arguments. Strings are always left
unchanged.  HTML terms may contain logic variables, provided they are
instantiated before the term is translated or output.  This allows
creating documents piecemeal, back-patching of references in documents,
etc.

In the following sections we list the meaning of the principal Prolog
structures that represent special functionality at the HTML level.  Only
special atoms are translated, the rest are assumed to be normal text and
will be passed through to the HTML document.

\subsection{General Structures}
\label{sec:genstruct}

Basically, HTML has two kinds of components: HTML {\em elements} and
HTML {\em environments}.  An HTML element has the form \mbox{``{\tt
 <NAME \var{Attributes} >}''} were {\tt NAME} is the name of the
element and \var{Attributes} is a (possibly empty) sequence of
attributes, each of them being either an attribute name or an attribute
assignment as \mbox{{\tt name="Value"}}.

An HTML environment has the form \mbox{``{\tt <NAME \var{Attributes} >
  \var{Text}}} \verb+</NAME>+'' were {\tt NAME} is the name of the
environment an \var{Attributes} has the same form as before.

The general Prolog structures that represent these two HTML 
constructions are:
\begin{descpredicates}
  \predicateitem{\var{Name}\,\$\var{Atts}} (`{\tt \$/2}' is defined as
  an infix, binary
  operator.) Represents an HTML element of name \var{Name} and
  attributes \var{Atts}, were \var{Atts} is a (possibly empty) list of
  attributes, each of them being either an atom or a structure {\tt
    \var{name}=\var{value}}. For example, the term
  \begin{quote}
    \verb+img$[src='images/map.gif',alt="A map",ismap]+ 
  \end{quote}
  is translated into the HTML source
  \begin{quote}
    \verb+<img src="images/map.gif" alt="A map" ismap>+
  \end{quote}
  Note that HTML is not case-sensitive, so we can use lower-case
  atoms.
  \predicateitem{\var{name}(\var{Text})} (A term with functor
  \var{name}/1 and argument \var{Text}) Represents an  HTML
  environment of name \var{name} and included text \var{Text}. For
  example, the term
  \begin{quote}
    \verb+address('clip@dia.fi.upm.es')+
  \end{quote}
  is translated into the HTML source
  \begin{quote}
    \verb+<address>clip@dia.fi.upm.es</address>+
  \end{quote}
  \predicateitem{\var{name}(\var{Atts},\var{Text})} (This is a term with
  functor \var{name}/2 and arguments \var{Atts} and \var{Text})
  Represents an  HTML environment of name \var{name}, attributes
  \var{Atts} and included text \var{Text}. For example, the term
  \begin{quote}
    \verb+a([href='http+\verb+://www.clip.dia.fi.upm.es/'],"Clip home")+
  \end{quote}
  represents the HTML source
  \begin{quote}
    \verb+<a href="http://www.clip.dia.fi.upm.es/">Clip home</a>+
  \end{quote}
  \predicateitem{env(\var{Name},\var{Atts},\var{Text})} Equivalent to
  \texttt{\var{Name}(\var{Atts},\var{Text})}.
  \predicateitem{begin(\var{Name},\var{Atts})} It translates to the start
  of an HTML environment of name \var{Name} and attributes \var{Atts}. There
  exists also a {\tt begin(\var{Name})} structure.  Useful, in conjunction
  with the next structure, when including in a document output generated
  by an existing piece of code (e.g.\ \var{Name} = {\tt pre}).  Its use is
  otherwise discouraged.
  \predicateitem{end(\var{Name})} Translates to the end of an HTML
  environment of name \var{Name}.
\end{descpredicates}

Now we can rewrite the previous example as follows (note how the use
of the logic variable \texttt{Response} allows injecting the result of
the call to \texttt{response/1} into the output term, using unification):
\begin{verbatim}
#!/usr/local/bin/ciao-shell

:- include(library(pillow)).

main(_) :-
    get_form_input(Input),
    get_form_value(Input,person_name,Name),
    response(Name,Response),
    output_html([
        'Content-type: text/html\n\n',
        html([title('Telephone database'),
              img$[src='phone.gif'],
              h2('Telephone database'),
              hr$[],
              Response])]). %% Using the logic variable.

response(Name, Response) :-
    form_empty_value(Name) ->
       Response = 'You have to provide a name.'
  ; phone(Name, Phone) ->
       Response = ['Telephone number of ',b(Name),': ',Phone]
  ; Response = ['No telephone number available for ',b(Name),'.'].

phone(daniel, '336-7448').
phone(manuel, '336-7435').
phone(sacha,  '543-5316').
\end{verbatim}

Any HTML construction can be represented with these structures (except
comments and declarations, which could be included as atoms or strings),
but the \wlib\ library provides additional, specific structures to
simplify HTML creation.

\subsection{Specific Structures}
\label{spestruct}

In this section we will list some special structures for
HTML which \wlib\ understands.
While in many cases using the general structures (with the native HTML
names) is probably good practice, using specific structures such as
these can sometimes be convenient. 
Also, some of these structures have special functionality (e.g.,
\texttt{prolog\_term/1}). 
A predicate {\tt html\_expansion/2} is provided which allows 
defining new structures (tables, layers, etc).
Specific structures include (the reader is referred to the \wlib\
manual for a full listing):

\begin{descpredicates}
\predicateitem{start} Used at the beginning of a document (translates to
\verb+<html>+).

\predicateitem{end} Used at the end of a document (translates to \verb+</html>+). 

\begin{latexonly}
\predicateitem{--} Produces a horizontal rule (translates to \verb+<hr>+).

\predicateitem{\tt\char'134\char'134} Produces a line break
    (translates to \verb+<br>+). 
\end{latexonly}
\begin{htmlonly}
\predicateitem{-{\tt -}} Produces a horizontal rule (translates to
\verb+<hr>+).

\predicateitem{\verb+\\+} Produces a line break (translates to \verb+<br>+). 
\end{htmlonly}

\predicateitem{\$} Produces a paragraph break (translates to \verb+<p>+).

\predicateitem{comment(\var{Comment})} Used to insert an HTML comment
(translates to \mbox{{\tt <!-- \var{Comment} -->}}).

\predicateitem{declare(\var{Decl})} Used to insert an HTML declaration
-- seldom used (translates to {{\tt <!\var{Decl}>}}).

\predicateitem{image(\var{Addr})} Used to include an image of address (URL)
\var{Addr} (translates to an \verb+<img>+ element).

\predicateitem{image(\var{Addr},\var{Atts})} As above with the list of
attributes \var{Atts}.

\predicateitem{ref(\var{Addr},\var{Text})} Produces a hypertext link, \var{Addr}
is the URL of the referenced resource, \var{Text} is the text of the
reference (translates to
\verb+<a href="+\var{Addr}\verb+">+\var{Text}\verb+</a>+).

\predicateitem{label(\var{Label},\var{Text})} Labels \var{Text} as a target
destination with label \var{Label} (translates to \verb+<a name="+%
\var{Label}\verb+">+\var{Text}\verb+</a>+).

\predicateitem{heading(\var{N},\var{Text})} Produces a heading of level
\var{N} ($1 \leq N \leq 6$), \var{Text} is the text to be used as
heading -- useful when one wants a heading level relative to
another heading (translates to a \verb+<h+\var{N}\verb+>+ environment).

\predicateitem{itemize(\var{Items})} Produces a list of bulleted items,
\var{Items} is a list of corresponding HTML terms (translates to a
\verb+<ul>+ environment).

\predicateitem{enumerate(\var{Items})} Produces a list of numbered items,
\var{Items} is a list of corresponding HTML terms (translates to an
\verb+<ol>+ environment).

\predicateitem{description(\var{Defs})} Produces a list of defined items,
\var{Defs} is a list whose elements are definitions, each of them
being a Prolog sequence (composed by {\tt ','}/2 operators). The last
element of the sequence is the definition, the other (if any) are the
defined terms (translates to an \verb+<dl>+ environment).

\predicateitem{nice\_itemize(\var{Img},\var{Items})} Produces a list of
bulleted items, using the image \var{Img} as bullet. The predicate {\tt
  icon\_address/2} provides a colored bullet.

\predicateitem{preformatted(\var{Text})} Used to include preformatted text,
\var{Text} is a list of HTML terms, each element of the list being a
line of the resulting document (translates to a \verb+<pre>+
environment).

\predicateitem{verbatim(\var{Text})} Used to include text verbatim, special
HTML characters ({\tt <,>,\&,"}) are translated into its quoted HTML 
equivalent.

\predicateitem{prolog\_term(\var{Term})} Includes any prolog term
   \var{Term}, represented in functional notation.  Variables are output
   as \texttt{\_}.

\predicateitem{nl} Used to include a newline in the HTML source (just to
improve human readability).

\predicateitem{entity(\var{Name})} Includes the entity of name
\var{Name} (ISO-8859-1 special character).

\predicateitem{cgi\_reply} This is not HTML, rather, the CGI protocol
requires this content descriptor to be used by CGI executables
(including form handlers) when replying (translates to ``{\tt
  Content-type: text/html}'').

\predicateitem{pr} Includes in the page a graphical logo with the
  message ``Developed using the \wlib\ Web programming library'', which
  points to the manual and library source.
\end{descpredicates}

With these additional structures, we can rewrite the previous example as
follows (note that in this example the
use of {\tt heading/2} or {\tt h2/1} is equally suitable):
\begin{verbatim}
#!/usr/local/bin/ciao-shell

:- include(library(pillow)).

main(_) :-
    get_form_input(Input),
    get_form_value(Input,person_name,Name),
    response(Name,Response),
    output_html([
        cgi_reply,
        start,
        title('Telephone database'),
        image('phone.gif'),
        heading(2,'Telephone database'),
        --,
        Response,
        end]).

response(Name, Response) :-
    form_empty_value(Name) ->
       Response = 'You have to provide a name.'
  ; phone(Name, Phone) ->
       Response = ['Telephone number of ',b(Name),': ',Phone]
  ; Response = ['No telephone number available for ',b(Name),'.'].

phone(daniel, '336-7448').
phone(manuel, '336-7435').
phone(sacha,  '543-5316').
\end{verbatim}

We have not included above the specific structures for creating
forms. They are included and explained in the following section.

\subsection{Specific Structures for Forms}
\label{formsstruct}

In this section we explain the structures which represent the various
elements related to forms:

\begin{descpredicates}
  \predicateitem{start\_form(\var{Addr}$[$,\var{Atts}$]$)} Specifies
  the beginning of a form. \var{Addr} is the address (URL) of the
  program that will handle the form, and \var{Atts} other attributes of the
  form, as the method used to invoke it. If \var{Atts} is not
  present the method defaults to POST. (Translates to \linebreak
  \texttt{<form action="\var{Addr}" \var{Atts} >}.)

  \predicateitem{start\_form} Specifies the beginning of a form without
  assigning address to the handler, so that the form handler will be the
  cgi-bin executable producing the form.

  \predicateitem{end\_form} Specifies the end of a form (translates to
  \verb+</form>+).

  \predicateitem{checkbox(\var{Name},\var{State})} Specifies an input of type
  \verb+checkbox+ with name \var{Name}, \var{State}=on if the checkbox is
  initially checked (translates to an \verb+<input>+ element).

  \predicateitem{radio(\var{Name},\var{Value},\var{Selected})} Specifies an
  input of type \verb+radio+ with name \var{Name} (several radio buttons
  which are interlocked must share their name), \var{Value} is the the
  value returned by the button, if \var{Selected}=\var{Value} the button
  is initially checked (translates to an \verb+<input>+ element).

  \predicateitem{input(\var{Type},\var{Atts})} Specifies an input of type
  \var{Type} with a list of attributes \var{Atts}.  Possible values
  of \var{Type} are \verb+text+, \verb+hidden+, \verb+submit+,
  \verb+reset+, \ldots (translates to an \verb+<input>+ element).

  \predicateitem{textinput(\var{Name},\var{Atts},\var{Text})} Specifies
  an input text area of name \var{Name}. \var{Text} provides the default
  text to be shown in the area, \var{Atts} a list of attributes
  (translates to a \verb+<textarea>+ environment).
  
  \predicateitem{option(\var{Name},\var{Val},\var{Options})} Specifies a
  simple option selector of name \var{Name}, \var{Options} is the list
  of available options and \var{Val} is the initial selected option (if
  \var{Val} is not in \var{Options} the first item is selected)
  (translates to a \verb+<select>+ environment).

  \predicateitem{menu(\var{Name},\var{Atts},\var{Items})} Specifies a menu of
  name \var{Name}, list of attributes \var{Atts} and list of options
  \var{Items}. The elements of the list \var{Items} are marked with the
  prefix operator `{\tt \$}' to indicate that they are selected
  (translates to a \verb+<select>+ environment).
\end{descpredicates}

For example, in order to generate a form suitable for sending input to
the previously described phone database handler one could execute the
following goal:

\begin{alltt}
    output_html([
        start,
        title('Telephone database'),
        heading(2,'Telephone database'),
        $,
        start_form('http{}://www.clip.dia.fi.upm.es/cgi-bin/phone_db.pl'),
        'Click here, enter name of clip member, and press Return:', 
\end{alltt}
\begin{verbatim}
        \\,
        input(text,[name=person_name,size=20]),
        end_form,
        end]).
\end{verbatim}

Of course, one could have also simply written directly the resulting
HTML document: 

\begin{verbatim}
<html>
<title>Telephone database</title>
<h2>Telephone database</h2>
<p>
<form method="POST"
 action="http://www.clip.dia.fi.upm.es/cgi-bin/phone_db.pl">
Click here, enter name of clip member, and press Return:
<br>
<input type="text" name="person_name" size="20">
</form>
</html>
\end{verbatim}

\section{Merging the Form Producer and the Handler}
\label{sec:merging}

An interesting practice when producing HTML forms and handlers is to
merge the operation of the form producer and the handler into the same
program.  The idea is to produce a generalized handler which receives
the form input, parses it, computes the answer, and produces a new
document which contains the answer to the input, as well as a new form.
A special case must be made for the first invocation, in which the input
would be empty, and then only the form should be generated. The
following is an example which merges the producer and the handler for
the phones database:\footnote{Notice that when only one text field exists
  in a form, the form can be submitted by simply pressing ``Return'' inside
  the text field.}

\begin{verbatim}
#!/usr/local/bin/ciao-shell

:- include(library(pillow)).

main(_) :-
    get_form_input(Input),
    get_form_value(Input,person_name,Name),
    response(Name,Response),
    output_html([
        cgi_reply,
        start,
        title('Telephone database'),
        image('phone.gif'),
        heading(2,'Telephone database'),
        --,
        Response,
        start_form,
        'Click here, enter name of clip member, and press Return:', 
        \\,
        input(text,[name=person_name,size=20]),
        end_form,
        end]).

response(Name, Response) :-
    form_empty_value(Name) ->
       Response = []
  ; phone(Name, Phone) ->
       Response = ['Telephone number of ',b(Name),': ',Phone,$]
  ; Response = ['No telephone number available for ',b(Name),'.',$].

phone(daniel, '336-7448').
phone(manuel, '336-7435').
phone(sacha,  '543-5316').
\end{verbatim}

This combination of the form producer and the handler allows producing
applications that give the impression of being interactive, even if
each step involves starting and running the handler to
completion.  Note that forms can contain fields which are not
displayed and are passed as input to the next invocation of the
handler. This allows passing state from one invocation of the handler
to the next one.

Finally, a note about testing and debugging CGI scripts: this is
unfortunately not as straightforward as it could be.  Useful
techniques include carefully checking permissions, looking at the data
logs of the server, replacing predicates such as \texttt{get\_form\_...}
with versions that print what is really being received, etc.

\section{Templates}
\label{sec:templates}

A problem in the previous programs is that the layout of the output page
is not easily configurable --it is hard-coded in the source and can
only be changed by modifying the program.  This is something that a
normal user (or even an expert programmer if the size of the program is
large) may not want to do.  In order to address this, \wlib\ provides a
facility for reading in ``HTML templates'' (also XML templates), and
converting them into a term format in which it is very natural to
manipulate them. An HTML template is a file which contains standard HTML
code, but in which ``slots'' can be defined and given an identifier by
means of a special tag.  These slots represent parts of the HTML code in
which other HTML code can be inserted. Once the HTML template is read by
\wlib, such slots appear as free logic variables in the corresponding
\wlib\ terms.  In this way, the user can define a layout with an HTML
editor of choice, taking care of marking the ``left out'' parts with
given names. These parts will then be filled appropriately by the
program.  The functionality associated with parsing such terms is
encapsulated in the following predicate:

\begin{descpredicates}
  
  \predicateitem{html\_template(\var{Chars}, \var{Terms}, \var{Dict})}
  Parses the string \var{Chars} as the contents of an HTML template and
  unifies \var{Terms} with the list of HTML terms comprised in the
  template, substituting occurrences of the special tag
  \texttt{<V>}\textit{name}\texttt{</V>} with prolog variables.
  \var{Dict} is instantiated to the dictionary of such substitutions, as
  a list of \textit{name}=\textit{Variable} pairs.

\end{descpredicates}

In the following example a template file called \texttt{TlfDB.html} is
assumed to hold the formatting of the output page, defining an HTML
variable called ``response'' which will be substituted by the response
of the CGI program.  Note that the predicate
\texttt{file\_to\_string/2} (defined in Ciao library
\texttt{file\_utils}) reads a file and returns in its second
argument the contents of the file as a list of character codes. Note
also that calling \texttt{html\_template/3} with the third argument
instantiated to \texttt{[response = Response]} has the effect of
instantiating the ``slot'' in \texttt{HTML\_terms} to the contents of
\texttt{Response} (this makes use of the fact that there is only one
slot on the template; normally, a call to \texttt{member/2} is used
to locate the appropriate \textit{name}=\textit{Variable} pair). 

\begin{verbatim}
#!/usr/local/bin/ciao-shell

:- include(library(pillow)).
:- use_module(library(file_utils)).

main(_) :-
    get_form_input(Input),
    get_form_value(Input,person_name,Name),
    response(Name,Response),
    file_to_string('TlfDB.html', Contents),
    html_template(Contents, HTML_terms, [response = Response]),
    output_html([cgi_reply|HTML_terms]).

response(Name, Response) :-
    form_empty_value(Name) ->
       Response = []
  ; phone(Name, Phone) ->
       Response = ['Telephone number of ',b(Name),': ',Phone,$]
  ; Response = ['No telephone number available for ',b(Name),'.',$].

phone(daniel, '336-7448').
phone(manuel, '336-7435').
phone(sacha,  '543-5316').
\end{verbatim}

An example of the contents of the template file could be:

\begin{verbatim}
<HTML><HEAD><TITLE>Telephone database</TITLE></HEAD>
<BODY background="bg.gif">
<IMG src="phone.gif">
<H2>Telephone database</H2>
<HR>
<V>response</V>
<FORM method="POST">
Click here, enter name of clip member, and press Return:<BR>
<INPUT type="text" name="person_name" size="20"></FORM>
</BODY>
</HTML>
\end{verbatim}

\section{Accessing WWW documents}

The facilities presented in the previous sections allow generating
HTML documents, including forms, and handling the input coming from
forms. In many applications such as search tools, content analyzers,
etc., it is also desirable to be able to access documents on the
Internet. Such access is generally accomplished through protocols such as
{\tt FTP} and {\tt HTTP} which are built on top of TCP/IP.  In LP/CLP
systems which have TCP/IP connectivity (i.e., a sockets/ports
interface) the required protocols can be easily coded in the source
language using such facilities and DCG parsers.  At present, only the
HTTP protocol is supported by \wlib.  As with HTML code, the library
uses an internal representation of Uniform Resource Locators (URLs), to
be able to manipulate them easily, and provides predicates which translate
between the internal representation
and the textual form. The facilities provided by \wlib\ for accessing
WWW documents include the following predicates:

\begin{descpredicates}

  \predicateitem{url\_info(\var{URL},\var{Info})} Translates a URL
  \var{URL} to an internal structure \var{Info} which details its
  various components and vice-versa. For now non-HTTP URLs make the
  predicate fail. E.g.\\
  {\tt url\_info('http://www.foo.com/bar/scooby.txt',Info)}\\
  gives {\tt Info = http('www.foo.com',80,"/bar/scooby.txt")},\\
  {\tt url\_info(URL, http('www.foo.com',2000,"/bar/scooby.txt")}\\
  gives {\tt URL = "http://www.foo.com:2000/bar/scooby.txt"} (a string).

  \predicateitem{url\_info\_relative(\var{URL},\var{BaseInfo},\var{Info})}
  Translates a relative URL \var{URL} which appears in the HTML page
  referred to by \var{BaseInfo} (given as an {\tt url\_info} structure)
  to a complete {\tt url\_info} structure \var{Info}. Absolute URLs are
  translated as with the previous predicate. E.g.\\
  {\tt url\_info\_relative("/guu/intro.html", http('www.foo.com',80,"/bar/scoob.html"), Info)}\\
  gives {\tt Info = http('www.foo.com',80,"/guu/intro.html")}\\
  {\tt url\_info\_relative("dadu.html", http('www.foo.com',80,"/bar/scoob.html"), Info)}\\
  gives {\tt Info = http('www.foo.com',80,"/bar/dadu.html")}.

  \predicateitem{url\_query(\var{Dic},\var{Args})} Translates a list of
  \var{attribute}=\var{value} pairs \var{Dic} (in the same form as the
  dictionary returned by {\tt get\_form\_input/1}) to a string
  \var{Args} for appending to a URL pointing to a form handler.

  \predicateitem{fetch\_url(\var{URL},\var{Request},\var{Response})}
  Fetches a document from the Internet. \var{URL} is the Uniform
  Resource Locator of the document, given as a {\tt url\_info} structure.
  \var{Request} is a list of options which specify the parameters of the
  request, \var{Response} is a list which includes the parameters of the
  response. The request parameters available include:
  \begin{description}
    \item[{\tt head}] To specify that we are only interested in the
      header.
    \item[{\tt timeout(\var{Time})}] \var{Time} specifies the maximum
      period of time (in seconds) to wait for a response. The
      predicate fails on timeout.
    \item[{\tt if\_modified\_since(\var{Date})}] Get document only if
      newer than \var{Date}. An example of a structure that
      represents a date is
      \verb+date('Tuesday',15,'January',1985,'06:14:02')+.
    \item[{\tt user\_agent(\var{Name})}] Provide a user-agent field.
    \item[{\tt authorization(\var{Scheme},\var{Params})}]
        Provides an authentication field when accessing restricted sites.
    \item[{\tt \var{name}(\var{Param})}] Any other functor translates
      to a field of the same name (e.g. \verb+from('user@machine')+).
  \end{description}
  The parameters which can be returned in the response list include (see the
      HTTP/1.0 definition for more information):
  \begin{description}
    \item[{\tt content(\var{Content})}] Returns in \var{Content} the
    actual document text, as a list of characters.

    \item[{\tt status(\var{Type},\var{Code},\var{Phrase})}] Gives the
      status of the response. \var{Type} can be any of
      \verb+informational+, \verb+success+, \verb+redirection+,
      \verb+request_error+, \verb+server_error+ or
      \verb+extension_code+, \var{Code} is the status code and
      \var{Phrase} is a textual explanation of the status.
    \item[{\tt pragma(\var{Data})}] Miscellaneous data.
    \item[{\tt message\_date(\var{Date})}]
        The time at which the message was sent. 
    \item[{\tt location(\var{URL})}] The document has moved to this URL.
    \item[{\tt http\_server(\var{Server})}] Identifies the server responding.
    \item[{\tt allow(\var{methods})}] List of methods allowed by the server.
    
    \item[{\tt last\_modified(\var{Date})}] Date/time at which the
      sender believes the resource was last modified.

    \item[{\tt expires(\var{Date})}] Date/time after which the entity
      should be considered stale.
    
    \item[{\tt content\_type(\var{Type},\var{Subtype},\var{Params})}]
        Returns the MIME type/subtype of the document.
    \item[{\tt content\_encoding(\var{Type})}] Encoding of the
      document (if any).

    \item[{\tt content\_length(\var{Length})}] \var{Length} is
        the size of the document, in bytes.

    \item[{\tt authenticate(\var{Challenges})}] Request for authentication.
  \end{description}

  \predicateitem{html2terms(\var{Chars},\var{Terms})} We have already
  explained how this predicate transforms HTML terms to HTML format.
  Used the other way around it can parse HTML code, for example retrieved by
  {\tt fetch\_url}. The resulting list of HTML terms \var{Terms} is
  normalized: it contains only {\tt comment/1}, {\tt declare/1}, {\tt
    env/3} and {\tt \$/2} structures.
      
\end{descpredicates}

For example, a simple fetch of a document can be done as follows:
\begin{alltt}
    url_info('http{}://www.foo.com',UI), fetch_url(UI,[],R),
    member(content(C),R), html2terms(C, HTML_Terms).
\end{alltt}
Note that if an error occurs (the document does not exist or has
moved, for example) this will simply fail. The following call
retrieves a document if it has been modified since October 6, 1999:
\begin{verbatim}
    fetch_url(http('www.foo.com',80,"/doc.html"),
        [if_modified_since('Wednesday',6,'October',1999,'00:00:00')],
        R).
\end{verbatim}
This last one retrieves the header of a document (with a timeout of 10
seconds) to get its last modified date:
\begin{verbatim}
    fetch_url(http('www.foo.com',80,"/last_news.html"),
              [head,timeout(10)],R),
    member(last_modified(Date),R).
\end{verbatim}

The following is a simple application illustrating the use of {\tt
  fetch\_url} and {\tt html2terms}. The example defines {\tt
  check\_links(\var{URL},\var{BadLinks})}. The predicate fetches the
HTML document pointed to by \var{URL} and scours it to check for links
which produce errors when followed.  The list
\var{BadLinks} contains all the bad links found, stored as compound
terms of the form: {\tt badlink(Link,Error)} where {\tt Link} is
the problematic link and {\tt Error} is the error explanation given by
the server.

\begin{verbatim}
check_links(URL,BadLinks) :-
        url_info(URL,URLInfo),
        fetch_url(URLInfo,[],Response),
        member(content_type(text,html,_),Response),
        member(content(Content),Response),
        html2terms(Content,Terms),
        check_source_links(Terms,URLInfo,[],BadLinks).

check_source_links([],_,BL,BL).
check_source_links([E|Es],BaseURL,BL0,BL) :-
        check_source_links1(E,BaseURL,BL0,BL1),
        check_source_links(Es,BaseURL,BL1,BL).

check_source_links1(env(a,AnchorAtts,_),BaseURL,BL0,BL) :-
        member((href=URL),AnchorAtts), !,
        check_link(URL,BaseURL,BL0,BL).
check_source_links1(env(_Name,_Atts,Env_html),BaseURL,BL0,BL) :- !,
        check_source_links(Env_html,BaseURL,BL0,BL).
check_source_links1(_,_,BL,BL).

check_link(URL,BaseURL,BL0,BL) :-
        url_info_relative(URL,BaseURL,URLInfo), !,
        fetch_url_status(URLInfo,Status,Phrase),
        ( Status \== success ->
          name(P,Phrase),
          name(U,URL),
          BL = [badlink(U,P)|BL0]
        ; BL = BL0
        ).
check_link(_,_,BL,BL).

fetch_url_status(URL,Status,Phrase) :-
        fetch_url(URL,[head,timeout(20)],Response), !,
        member(status(Status,_,Phrase),Response).
fetch_url_status(_,timeout,"Timeout").
\end{verbatim}

\section{Providing Code Through the WWW}

A facility which can be easily built on top of the primitives presented
so far is that of ``remote WWW modules,'' i.e., program modules which
reside on the net at a particular HTTP address in the same way that
normal program modules reside in a particular location in the local file
system. This allows for example always fetching the most recent version
of a given library (e.g., \wlib) when a program is compiled.
For example, the form handler of Section \ref{sec:genstruct}, if rewritten
as
\begin{alltt}
#!/usr/local/bin/ciao-shell

:- use_module('http{}://www.clip.dia.fi.upm.es/lib/pillow.pl').

main(_) :-
    get_form_input(Input),
    get_form_value(Input,person_name,Name),
...
\end{alltt}
would load the current version of the library each time it is executed.
This generalized module declaration is just syntactic sugar, using
{\tt expand\_term}, for a document fetch, using {\tt fetch\_url},
followed by a standard {\tt use\_module} declaration.  It is obviously
interesting to combine this facility with caching strategies. 
An interesting (and straightforward to implement) additional feature
is to fetch remote byte-code (as generally done by {\tt use\_module}),
if available, but this is only possible if the two systems use the
same byte-code (this can normally be checked easily in the bytecode itself). 
Also, it may be interesting to combine this type of code downloading
with WWW document accesses, so that code is downloaded automatically
when a particular document is fetched. 
This issue is addressed in Section \ref{auto-download}. 
Finally, there are obvious security issues related to downloading code
in general, 
which can be addressed with standard techniques such as security 
signatures.

\section{A High-Level Model of Client-Server Interaction: Active Modules}

Despite its power, the cgi-bin interface also has some shortcomings.
The most serious is perhaps the fact that the handler is started and
expected to terminate for each interaction. This has two
disadvantages. First, no state is preserved from one query to the
next. However, as mentioned before, this can be fixed by passing the
state through the form (using \emph{hidden} fields), by saving it in a
temporary file at the server side, by using ``cookies'', etc. Second,
and more importantly, starting and stopping the application may be
inefficient. For example, if the idea is to query a large database or
a natural language understanding system, it may take a long time to
start and stop the system. In order to avoid this we propose an
alternative architecture for cgi-bin applications (a similar idea,
although not based on the idea of active modules, has been proposed
independently by Ken Bowen \cite{ALS-html-pl-www}).

\begin{figure}
\centerline{\includegraphics{\figpath/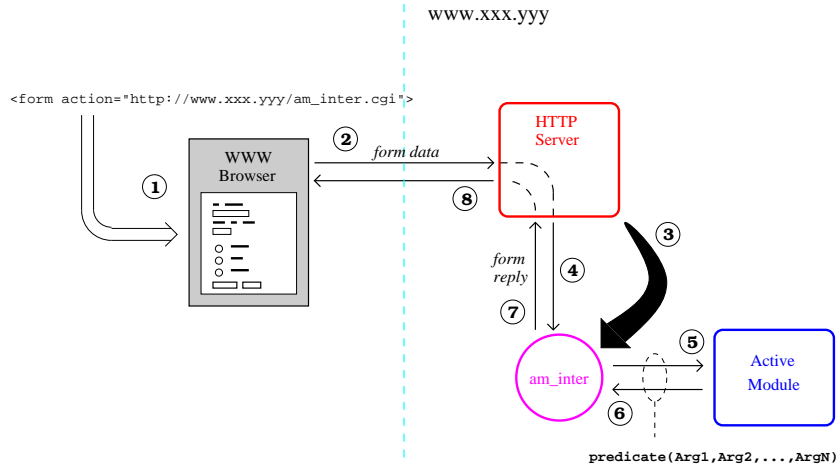}}
\caption{The Forms Interface using Active Modules}
\label{fig:formactive}
\end{figure}

The basic idea is illustrated in Figure \ref{fig:formactive}. The
operation is identical to that of standard form handlers, as
illustrated in Figure \ref{fig:forms}, up to step 3. In this step, the
handler started is not the application itself, but rather an interface
to the actual application, which is running continuously and thus
contains state. Thus, only the interface is started and stopped with
every transaction. The interface simply passes the form input received
from the server (4) to the running application (5) and then forwards
the output from the application (6) to the server before terminating,
while the application itself continues running. Both the interface and
the application can be written in LP/CLP, using the predicates
presented. The interface can be a simple script, while the application
itself will be typically compiled. 

An interesting issue is that of communication between interface and
application. This can of course be done through sockets. However, as a
cleaner and much simpler alternative, the concept of active modules
\cite{ciao-dis-impl-parimp-www} can be used to advantage in this
application.  An active module (or an active object, if modularity is
implemented via objects) is an ordinary module to which computational
resources are attached (for example, a process on a UNIX machine), and
which resides at a given (socket) address on the network.\footnote{It
  is also possible to provide active modules via a WWW address.
  However, we find it more straightforward to simply use socket
  addresses. In any case, this is generally hidden inside the access
  method and can be thus made transparent to the user.} Compiling an
active module produces an executable which, when running, acts as a
server for a number of relations, which are the predicates exported by
the module. The relations exported by the active module can be
accessed by any program on the network by simply ``loading'' the
module and thus importing such ``remote relations.''  The idea is that
the process of loading an active module does not involve transferring
any code, but rather setting up things so that calls in the local
module are executed as remote procedure calls to the active module,
possibly over the network.  Except for compiling it in a special way, an
active module is identical from the programmer point of view to an
ordinary module. Also, a program using an active module imports it and
uses it in the same way as any other module, except that it uses
``{\tt use\_active\_module}'' rather than ``{\tt use\_module}'' (see
below). Also, an active module has an address (network address) which
must be known in order to use it.  The address can be announced by the
active module when it is started via a file or a name server (which
would be itself another active module with a fixed address).

We now present the constructs related to active modules in Ciao:
\begin{descpredicates}
\predicateitem{:- use\_active\_module(\var{Module},\var{Predicates})}
  A declaration used to import the predicates in the list
  \ttvar{Predicates} from the active module \ttvar{Module}. From this point
  on, the code should be written as if a standard {\tt use\_module}/2
  declaration had been used.  The declaration needs the following
  predicate to be accessible from the module.
\predicateitem{module\_address(\var{Module},\var{Address})}
  This predicate must return in \ttvar{Address} the address of
  \ttvar{Module}, for any active module imported in the code.  There are a
  number of standard libraries defining versions of this predicate.
\predicateitem{save\_addr\_actmod(\var{Address})}
  This predicate should define a way to publish \ttvar{Address},
  to be used in active modules (the name of the active module is taken
  as the name of the current executable).  There are a number of standard
  libraries defining versions of this predicate, which are in
  correspondence with the libraries which define versions of the
  previous predicate.
\predicateitem{make\_actmod(\var{ModuleFile},\var{PublishModule})}
  Makes an active module executable from the module residing in
  \ttvar{ModuleFile}, using address publish module of name
  \ttvar{PublishModule}.  When the executable is run (for example, at the
  operating system level by ``\ttvar{Module} {\tt \&}''), a socket is
  created and the hook predicate 
  \texttt{save\_addr\_actmod/1} mentioned above (which is
  supposed to be defined in \ttvar{PublishModule}) is called in order to
  export the active module address as required.  Then, a standard
  driver is run to attend network requests for the module exported
  predicates.  Note that the code of \ttvar{ModuleFile} does not need to
  be written in any special way.

\end{descpredicates}

This scheme is very flexible, allowing to completely configure the way
active modules are located.  This is accomplished by writing a pair of
libraries, one defining the way an active module address is published,
and a second defining the way the address of a given active module is
found.  For example, the Ciao standard libraries include as an example
an implementation (libraries \texttt{filebased\_publish} and
\texttt{filebased\_locate}) which uses a directory accessible by all
the involved machines (via NFS) to store the addresses of the active
modules, and the \texttt{module\_address}/2 predicate examines this
directory to find the required data. Other solutions provided as
examples include posting the address at a WWW address
(\texttt{webbased\_publish} / \texttt{webbased\_locate}), and an
implementation of a name server, that is, another active module 
(this one with a known, fixed address) that records the addresses of
active modules and supplies this data to the modules that import it,
serving as a contact agency between servers and clients.

From the implementation point of view, active modules are essentially
daemons: Prolog executables which are started as independent processes
at the operating system level.  In the Ciao system library,
communication with active modules is implemented using sockets (thus,
the address of an active module is a UNIX socket in a machine).
Requests to execute goals in the module are sent through the socket by
remote programs.  When such a request arrives, the process running the
active module takes it and executes it, returning through the socket
the computed results. These results are then taken by the remote
processes.

Thus, when the compiler finds a {\tt use\_active\_module} declaration,
it defines the imported predicates as remote calls to the active module.
For example, if the predicate \texttt{\textit{P}} is imported from the active
module \texttt{\textit{M}}, the predicate would be defined as
\begin{quote}
  {\ttfamily
  \textit{P} :- module\_address(\textit{M},A), remote\_call(A,\textit{P})  }
\end{quote}

Compiling the following code as an active module, by writing at the
Ciao toplevel ``\texttt{make\_actmod(phone\_db,
  'actmods/filebased\_publish')}'' (or, using the stand\-alone compiler, by
executing ``\texttt{ciaoc -a 'actmods/filebased\_publish' phone\_db}''),
creates an executable {\tt phone\_db} which, when started as a process
(for example, by typing ``{\tt phone\_db \&}'' at a UNIX shell prompt)
saves its address (i.e., that of its socket) in file
\texttt{phone\_db.addr} and waits for queries from any module which
``imports'' this module (it also provides a predicate to dynamically add
information to the database):

\begin{verbatim}
:- module(phone_db,[response/2,add_phone/2]).

response(Name, Response) :-
    form_empty_value(Name) ->
       Response = 'You have to provide a name.'
  ; phone(Name, Phone) ->
       Response = ['Telephone number of ',b(Name),': ',Phone]
  ; Response = ['No telephone number available for ',b(Name),'.'].

add_phone(Name, Phone) :-
    assert(phone(Name, Phone)).

:- dynamic phone/2.
phone(daniel, '336-7448').
phone(manuel, '336-7435').
phone(sacha,  '543-5316').

\end{verbatim}

The following simple script can be used as a cgi-bin executable which
will be the active module interface for the previous active module.
When started, it will process the form input, issue a call to {\tt
  response/2} (which will be automatically handled by the {\tt
  phone\_db} active module), and produce a new form before
terminating. It will locate the address of the {\tt phone\_db} active
module via the {\tt module\_address/2} predicate defined in library
\texttt{'actmods/filebased\_locate'}.

\begin{verbatim}
#!/usr/local/bin/ciao-shell

:- use_active_module(phone_db,[response/2]).
:- use_module(library('actmods/filebased_locate')).
:- include(library(pillow)).


main(_) :-
    get_form_input(Input),
    get_form_value(Input,person_name,Name),
    response(Name,Response),
    output_html([
        cgi_reply,
        start,
        title('Telephone database'),
        image('phone.gif'),
        heading(2,'Telephone database'),
        --,
        Response,
        $,
        start_form,
        'Click here, enter name of clip member, and press Return:', 
        \\,
        input(text,[name=person_name,size=20]),
        end_form,
        end]).
\end{verbatim}

There are many enhancements to this simple schema which, for brevity,
are only sketched here. One is to add concurrency to the active module
(or whatever means of handling the client-server interaction is being
used), in order to handle queries from different clients concurrently.
This is easy to do in systems that support concurrency natively, such
as Ciao, BinProlog/$\mu^2$-Prolog, AKL, Oz, and KL1.  We feel that
Ciao can offer advantages in this area because it offers compatibility
with Prolog and CLP systems while at the same time efficiently
supporting concurrent execution of clause goals via local or
distributed threads~\cite{shared-database}.  Such goals can
communicate at different levels of abstraction: sockets/ports, the
shared fact database (similarly to a blackboard), or shared variables.
BinProlog/$\mu^2$-Prolog also supports threads, with somewhat
different communication
mechanisms~\cite{binprolog-complangprolog-www,multi-prolog}.  Finally,
as shown in \cite{aurora-www}, it is also possible to exploit the
concurrency present in or-parallel Prolog systems such as Aurora for
implementing a multitasking server.

It is also interesting to set up things so that a single active module
can handle different forms.  This can be done even dynamically (i.e.,
the capabilities of the active module are augmented on the fly, being
able to handle a new form), by designating a directory in which code
to be loaded by the active module would be put, the active module
consulting the directory periodically to increase its functionalities.
Finally, another important issue that has not been addressed is that
of providing security, i.e., ensuring that only allowed clients
connect to the active module. As in the case of remote code
downloading, standard forms of authentication based on codes can be
used.

\section{Automatic Code Downloading and Local Execution}
\label{auto-download}

\begin{figure}
\centerline{\includegraphics{\figpath/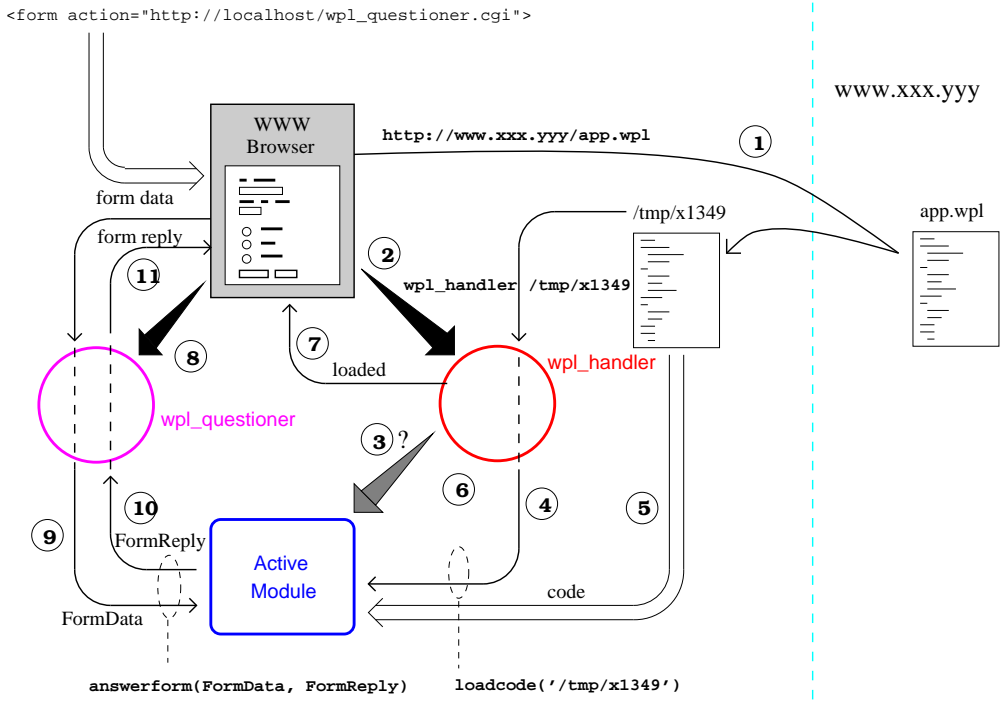}}
\caption{Automatic code downloading architecture}
\label{fig:prologweb}
\end{figure}

In this section we describe an architecture which, using only the
facilities we have presented in previous sections, allows the
downloading and local execution of Prolog (or other LP/CLP) code by
accessing a WWW address, without requiring a special browser.  This is a
complementary approach to giving WWW access to an active module in the
sense that it provides code which will be executed in the client machine
(\`{a} la Java). More concretely, the functionality that we desire is
that by simply clicking on a WWW pointer, and transparently for the
user, remote Prolog code is automatically downloaded in such a way that
it can be queried via forms and all the processing is done locally.

To allow this, the HTTP server on the server machine is 
configured to give a specific {\tt mime.type} (for example {\tt
  application/x-prolog}) to the files which will hold WWW-downloadable
Prolog code (for example those with a special suffix, like {\tt
  .wpl}).  On the other side, the browser is configured to
start the {\tt wpl\_handler} helper application when receiving data of
type {\tt application/x-prolog}. This {\tt wpl\_handler} application is
the interface to a Prolog engine which will execute the WWW downloaded
code, acting as an active module. We now sketch the procedure (see
figure \ref{fig:prologweb}):
\begin{enumerate}
\item The form that will be used to query the downloaded code (and
  which we assume already loaded on the browser) contains a link which
  points to a WWW-aware Prolog code file. Clicking on this link
  produces the download as explained below. Note that for browsers that can
  handle {\tt multipart/mixed} mime types (such as most modern
  browsers), the form and the code file 
  could alternatively be combined in the same document. However, for
  brevity, we will only describe the case when they are separate. The
  handler for 
  the form is specified as the local cgi-bin executable {\tt
    wpl\_questioner.cgi}. 
\item As the server of the file tells the browser that this page is of
  type {\tt application/x-prolog}, the browser starts a {\tt wpl\_handler} and
  passes the file to it (in this example by saving the file in a
  temporal directory and passing its name).
\item The {\tt wpl\_handler} process checks whether a Prolog engine is
  currently running for this browser and, if necessary, starts one. This
  Prolog engine is configured as an active module.
\item Then, through a call to a predicate of the active module
  ``{\tt loadcode(\var{File})}'' the handler asks
  the active module to read the code.
\item The active module reads the code and compiles it.
\item {\tt wpl\_handler} waits for the active module to complete the
  compilation and writes a ``done'' message to the browser. 
\item The browser receives the ``done'' message. 
\item\label{anotherquestion} Now, when the ``submit'' button in the
  form is pressed, and following the standard procedure for forms, the
  browser starts a {\tt
    wpl\_questioner} process, sending it the form data.
\item The {\tt wpl\_questioner} process gets this form data,
  translates it to a dictionary \var{FormData} and passes it to the
  active module through a call to its exported predicate {\tt
    answerform(\var{FormData},\var{FormReply})}. 
\item The active module processes this request, and returns in
  \var{FormReply} a WWW page (as a term) which contains the
  answer to it (and possibly a new form).
\item The {\tt wpl\_questioner} process translates \var{FormReply} to
  raw HTML and gives it back to the browser, dying afterwards.
  Subsequent queries to the active module can be accomplished either by
  going back to the previous page (using the ``back'' button present in
  many browsers) or, if the answer page contains a new query form, by
  using it. In any case, the procedure continues at \ref{anotherquestion}.
\end{enumerate}

The net effect of the approach is that by simply clicking on a WWW
pointer, remote Prolog code is automatically downloaded to a local
Prolog engine.  Queries posed via the form are answered locally by the
Prolog engine. 

There are obvious security issues that need to be taken care of in
this architecture. Again, standard authentication techniques can be
used. However, 
since source code is being passed around, 
it is comparatively easy to verify that no dangerous predicates (for
example, perhaps those that can access files) are executed. Note again
that it is also possible to download bytecode, since this is supported by
most current LP/CLP systems, using a similar approach.

\section{Related Work}

Previous general purpose work on WWW programming using computational
logic systems includes, to the best of our knowledge, the publicly
available {\tt html.pl} library \cite{html-pl-www} and manual, and the
LogicWeb system \cite{logicweb-www} (the \wlib\ library was also
described previously in \cite{pillow-ws}).  The {\tt html.pl} library
was built by D.\ Cabeza and M.\ Hermenegildo, using input from L.\ 
Naish's forms code for NU-Prolog and M.\ Hermenegildo and F.\ Bueno's
experiments building a WWW interface to the CHAT-80 \cite{chat}
program. It was released as a publicly available WWW library for
LP/CLP systems and announced, among other places, in the Internet {\tt
  comp.lang.prolog} newsgroup \cite{html-pl-complangprolog-www}. The
library has since been ported to a large number of systems and adapted
by several Prolog vendors, as well as used by different programmers in
various institutions. In particular, Ken Bowen has ported the library
to ALS Prolog and extended it to provide group processing of forms and
an alternative to our use of active modules \cite{ALS-html-pl-www}.
The present work is essentially a significant extension of the {\tt
  html.pl} library.

The main other previous body of work related to general-purpose
interfacing of logic programming and the WWW that we have knowledge of
is the LogicWeb \cite{logicweb-www} system, by S.W. Loke and A.
Davison.  The aim of LogicWeb is to use logic programming to extend
the concept of WWW pages, incorporating in them 
programmable behavior and state. In this, it shares goals with Java. 
It also offers rich primitives for accessing code in remote pages and
module structuring.  The aims of LogicWeb are different from those of
{\tt html.pl}/\wlib.  LogicWeb is presented as a system itself, and
its implementation is done through a tight integration with the Mosaic
browser, making use of special features of this browser.  In contrast,
{\tt html.pl}/\wlib\ is a general purpose library, meant to be used by
a general computational logic systems and is browser-independent. {\tt
  html.pl}/\wlib\ offers a wide range of functionalities, such as
syntax conversion between HTML and logic terms, access 
predicates for WWW pages, predicates for handling forms, etc., which
are generally at a somewhat lower level of abstraction than those of
LogicWeb. We believe that using \wlib\ and the ideas sketched in this
paper it is possible to add the quite interesting
functionality offered by LogicWeb to standard LP and CLP systems.  
We have shown some examples including access to passive remote
code (modules with an {\tt ftp} or {\tt http} address) from programs
and automatic remote code access and querying using standard browsers
and forms. In addition, we have discussed active remote code, where
the functionality, rather than the code itself, is exported.

More recently, a larger body of work on the topic was presented at the
workshop held on the topic of Logic Programming and the Internet at
the 1996 Joint International Conference and Symposium on Logic
Programming (where also a previous version of this paper was
presented).  The work presented in \cite{lightweb-www} is based on
LogicWeb, and aims to provide distributed lightweight databases on the
WWW. As with the basic LogicWeb system, we believe that the \wlib\ 
library can be used to implement in other systems the interesting
ideas proposed therein.  As briefly mentioned before, the work in
\cite{aurora-www} proposes an architecture similar to that of our
active modules in order to handle form requests. In this solution the
handling multiple requests is performed by using or-parallelism.
While we feel that and-parallelism (as in \&-Prolog's or Ciao's
threads) is more natural for modeling this kind of concurrency, the
ideas proposed are quite interesting.  The ECLiPSe HTTP-library
\cite{eclipse-agents-www}, aimed at implementing INTERNET agents,
offers functionality that is in part similar to that of the Ciao {\tt
  html.pl}/\wlib\ libraries, including facilities that are similar to
our active modules. The approach is different, however, in several
respects. The ECLiPSe library implements special HTTP servers and
clients. In contrast, \wlib\ uses standard HTTP servers and
interfaces.  Using special purpose servers may be interesting because
the approach possibly allows greater functionality. On the other hand
this approach in general requires either the substitution of the
standard server on a given machine or setting the special server at a
different socket address from the standard one. The ECLiPSe library
also contains functionality that is related to our active modules,
although the interface provided is at a lower level.  Finally, other
papers describing very interesting WWW applications are being
presented regularly, which underline the suitability of computational
logic systems for the task.  We believe that the Ciao \wlib\ library
can contribute to making it even easier to develop such applications
in the future.

Additional work on the topic of Logic Programming and the Internet can
be found in the proceedings of the workshop sponsored by the
Compulog-Net research network. The reader is referred to the tutorials
and papers presented in these two workshops for more information on a
number of applications, other libraries, and topics such as
interfacing and compilation from computational logic systems to Java.
Examples of Prolog systems interfaced with Java are BinProlog (see
\htmladdnormallink{{\tt
    http://clement.info.umoncton.ca/BinProlog}}{http://clement.info.umoncton.ca/BinProlog}),
Ciao~\cite{ciao-reference-manual-tr}, and
others~\cite{InterProlog-PACLP-Keynote}. 
Experimental Prolog to Java compilers have been built both in academia
(see for example jProlog at \htmladdnormallink{{\tt
    http://www.cs.kuleuven.ac.be/\~{}bmd/PrologInJava/}}{http://www.cs.kuleuven.ac.be/~bmd/PrologInJava/})
and Commercially (see for example the IF Prolog tools
\htmladdnormallink{{\tt
    http://www.ifcomputer.com}}{http://www.ifcomputer.com}).  This
approach is quite attractive, although the results cannot compete in
performance with conventional Prolog compilers (it is open for
research whether improvements in Java performance or improved
Prolog-to-Java compilation technology can bridge the gap). Other
commercial work on the topic of interfacing Prolog and the WWW (in
addition to that done on the ALS system mentioned above) include the
Amzi! Prolog WebLS System (\htmladdnormallink{{\tt
    http://www.amzi.com/share.htm }}{http://www.amzi.com/share.htm})
and the LPA PrologWeb System (\htmladdnormallink{{\tt
    http://www.lpa.co.uk }}{http://www.lpa.co.uk}).

Recent work using \wlib\ includes the ``Web
Integrator''\cite{webbases} --a webbase system that integrates data
from various Web sources, and allows users to query these Web sources
as if they were a single database-- and \texttt{WebDB} \cite{webdb}
--a WWW-based database management interface. Also, within the RadioWeb
project~\cite{radioweb-ta}, we have developed (in collaboration with
the group of M. Codish at Ben Gurion University) a constraint-based
language for describing WWW page layout and style rules and an engine
which, by interpreting these rules, can generate WWW sites which
dynamically adapt to parameters such as user
characteristics~\cite{radioweb-D2.2.M3}.

Additional applications developed with the \wlib\ library can be
accessed from the \wlib\ WWW site (see later).  A page with pointers
to the proceedings of the previously mentioned workshops, as well as
other information (including technical reports and tutorial) regarding
the topic of Logic Programming, Constraint Programming, and the
Internet is maintained at \htmladdnormallink{{\tt
    http://www.clip.dia.fi.upm.es/lpnet/}}{http://www.clip.dia.fi.upm.es/lpnet/}.

\section{Conclusions and Future Work}

We have discussed from a practical point of view a number of issues
involved in writing Internet and WWW applications using LP/CLP
systems.  In doing so, we have described \wlib, an Internet/WWW
programming library for LP/CLP systems. \wlib\ provides facilities for
generating HTML/XML structured documents, producing HTML forms,
writing form handlers, processing HTML/XML templates, accessing and
parsing WWW documents, and accessing code posted at HTTP addresses. We
have also described the architecture of some application classes,
including automatic code downloading, using a high-level model of
client-server interaction, {\em active modules}.  Finally we have also
described an architecture for automatic LP/CLP code downloading for
local execution, using generic browsers.  We believe that the Ciao
\wlib\ library can ease substantially the process of developing WWW
applications using computational logic systems.

We have recently developed several extensions to the library (for
example, for setting and getting ``cookies''), and sample applications
which make extensive use of concurrency (on those LP/CLP systems that
support it) to overlap network requests. We have also developed a
complementary library for interfacing Prolog with the Virtual Reality
Modeling Language VRML~\cite{provrml-paclp}.

In addition to being included as part of the Ciao system, the \wlib\ 
library is provided as a standard, standalone public domain library for
SICStus Prolog and other Prolog and CLP systems, supporting most of
its functionality.  Please contact the authors or consult our WWW site
\htmladdnormallink{\texttt{http://www.clip.dia.fi.upm.es}}{http://www.clip.dia.fi.upm.es}
and the \wlib\ page at
\htmladdnormallink{\texttt{http://www.clip.dia.fi.upm.es/\-Software/pillow/\-pillow.html}}{http://www.clip.dia.fi.upm.es/Software/pillow/pillow.html}
for download details and an up-to-date online version of the \wlib\ manual. The Ciao Prolog system is also freely available from 
\htmladdnormallink{\texttt{http://www.clip.dia.fi.upm.es}}{http://www.clip.dia.fi.upm.es}
and 
\htmladdnormallink{\texttt{http://www.ciaoprolog.org}}{http://www.ciaoprolog.org}.

\section*{Acknowledgments}

The authors would like to thank Lee Naish, Mats Carlsson, Tony
Beaumont, Ken Bowen, Michael Codish, Markus Fromherz, Paul Tarau,
Andrew Davison, and Koen De Bosschere for useful feedback on previous
versions of this document and the \wlib\ code.  The first versions of
the Ciao system and the {\tt html.pl} library were developed under
partial support from the ACCLAIM ESPRIT project. Subsequent
development has occurred in the context of MCYT projects ``ELLA''
and ``EDIPIA'' (MCYT TIC99-1151), ESPRIT project RADIOWEB, and
NSF/CICYT collaboration ``ECCOSIC'' (Fulbright 98059).

\end{document}